\documentclass{article}
\usepackage[utf8]{inputenc}
\usepackage[letterpaper,top=2cm,bottom=2cm,left=2.5cm,right=2.5cm,marginparwidth=1.75cm]{geometry}
\usepackage{amsfonts}
\usepackage{amsbsy}
\usepackage{amsmath}
\usepackage[numbers,sort]{natbib}
\usepackage{graphicx}
\usepackage{changepage}
\usepackage{caption}
\usepackage{makecell}
\usepackage{color, colortbl}

\usepackage{multirow}
\usepackage{algorithm} 
\usepackage{algpseudocode} 
\usepackage{lineno}
\usepackage{titling}
\usepackage{adjustbox}
\usepackage{url}

\title{Estimating uterine activity from electrohysterogram measurements via statistical tensor decomposition}

\author{Uri Goldsztejn$^1$ and Arye Nehorai$^{2,*}$}
\date{\small
    $^1$Department of Biomedical Engineering, McKelvey School of Engineering,\\ Washington University in St. Louis,  St. Louis, MO, USA\\%
    $^2$Preston M. Green Department of Electrical and Systems Engineering, McKelvey School of Engineering, Washington University in St. Louis, St. Louis, MO, USA\\
    $^{*}$ Corresponding author. Email address: nehorai@wustl.edu\\[2ex]%
}

\begin{document}

\maketitle

\section*{Abstract}

Complications during pregnancy and labor are common and can be especially detrimental in populations with limited access to healthcare. A promising technology to address these complications is the electrohysterogram (EHG), which measures abdominal electric potentials. Since EHG recordings are measured noninvasively, they record uterine electrical activity together with activity from other interfering sources, making their analysis more challenging. To facilitate the analysis of EHGs, we separate these measurements into uterine activity that is more variable across different electrodes and over time, which we term localized activity, and activity that is more evenly distributed in space and time. We represent multi-electrode EHGs as tensors and develop a Bayesian tensor decomposition for estimating localized and distributed electrical activities. To demonstrate that our method can estimate localized and distributed activities more accurately than existing methods, we simulate EHG measurements based on previous characterizations of the activities comprising these measurements. Furthermore, we demonstrate the effectiveness of our method in separating EHG bursts from other interfering activities recorded in EHGs using real measurements from two public datasets. Our results show that our method denoises EHG bursts with higher signal-to-noise ratios, defined as power ratios between EHG bursts and segments with only baseline EHG activity, than alternative methods. Such accurate estimation of uterine and other activities from EHG measurements could be useful for more precisely characterizing patterns of activity in EHGs, which is a step forward towards developing reliable, portable devices to reduce risks during pregnancy and labor.

\section*{Keywords}
Electrohysterogram recordings, statistical signal processing, tensor decomposition.

\section*{Introduction}

Complications during pregnancy and labor are especially frequent in populations with limited access to healthcare \cite{hillemeier2007individual,elixhauser2011complicating}.
Every year in the United States, about 900 mothers die from complications related to pregnancy and childbirth and approximately 50,000 mothers suffer from severe maternal morbidity \cite{cdc_maternal_mortality,centers2017severe}. To mitigate these complications, several technologies for monitoring uterine activity have been proposed in recent decades. For example, tocography uses an external tocodynamometer to measure the mechanical activity associated with uterine contractions and is useful for monitoring labor progression and fetal well-being \cite{schneider2014s1}. Although this mechanical activity can be measured more accurately by intrauterine pressure catheters (IUPC), IUPCs are invasive and increase the risk of complications, including uterine perforation and infections \cite{gee2020intrauterine,harper2013risks,rood2012complications}.

Various technologies have been developed to measure uterine electrical activity noninvasively. For example, magnetomyography measures the magnetic fields associated with uterine contractions, and it has been shown to predict imminent labor and to be useful for tracking magnetic activity changes towards labor \cite{eswaran2004prediction,govindan2015tracking}. However, magnetomyography is a costly technology that is not usually available in hospitals because it requires an electromagnetically shielded room and SQUID sensors \cite{eswaran2002first}.
Electromyometrial imaging (EMMI) combines magnetic resonance imaging (MRI) and a large array of electrodes to reconstruct electrical maps of uterine activity \cite{wu2019noninvasive}. Although this technology can accurately map uterine electrical activity, the complexity of this system limits its widespread use.
To estimate uterine electrical activity using EMMI, a pregnant mother would first need to undergo an MRI scan while wearing hundreds of MRI markers positioned around the abdomen and lower back, which can be costly and time-consuming. Afterwards, hundreds of electrodes need to be positioned in the locations of the MRI markers. Finally, the electrical activity measured from these electrodes needs to be combined with a mother-specific anatomical model generated from the MRI scan to estimate uterine electrical maps. As Wu et al.\ have noted, EMMI would need to be significantly simplified to be used as a clinical tool \cite{wu2019noninvasive}.

Electrohysterogram (EHG) recordings measure abdominal electric potentials and can be implemented in portable devices. EHG recordings have been used to develop practical applications, including predicting premature births and estimating intrauterine pressure during contractions  \cite{xu2022review,rabotti2008estimation,rooijakkers2014low,mischi2017dedicated,hassan2012better,goldsztejn2022predicting}.

Applications based on EHG measurements generally include preprocessing steps to isolate measurement components of interest \cite{xu2022review,hassan2011combination}. The uterine activity measured in EHGs is composed of the slow wave (SW) and the fast wave, which is further subdivided into the fast wave low (FWL) and the fast wave high (FWH) \cite{batista2016multichannel,xu2022review,terrien2007spectral}. The SW has spectral power between 0.01 and 0.03 Hz and higher amplitudes than the fast wave \cite{xu2022review,batista2016multichannel,garfield2021review}. The amplitude of the SW has been reported to be generally between 1 and 15 mV, whereas the amplitude of the fast wave is usually under 1 mV \cite{garfield2021review}.
The spectral power of the FWL is located between 0.1 and 0.3 Hz, while the spectrum of the FWH is located between 0.3 and 0.9 Hz \cite{terrien2007spectral,batista2016multichannel}. Besides uterine electrical activity, EHGs record activity resulting from multiple other sources, including: maternal respiration, between 0.2 and 0.34 Hz; maternal cardiac activity, around 1.2 to 1.5 Hz; motion artifacts, which increase the spectral energy between 1 and 4 Hz and also modify the energy between 0.1 and 1 Hz; activity of the abdominal muscles, around 30 Hz; and power line noise, that may be present at 50 or 60 Hz \cite{batista2016multichannel,hassan2011combination,xu2022review,ye2014automatic}. Notably, the spectra of EHG components can vary among different mothers, gestational stage, electrode configuration used to record the EHG, and the method used for estimating the power spectral density \cite{aydin2009comparison,rooijakkers2014influence,batista2016multichannel,terrien2010synchronization,garfield2021review}. Consequently, different frequency ranges and amplitudes have been reported for the EHG components \cite{garfield2021review,devedeux1993uterine}. 
Measurement components of interest, which depend on the intended application, are usually extracted using spectral filtering and ground or reference electrodes \cite{xu2022review,jager2018characterization}. Spectral filtering helpfully removes activity with spectral components that do not overlap with EHG activity, e.g., abdominal muscle activity and power line noise. However, spectral filtering is not sufficient for denoising EHG activity, because if spectral filtering is used to remove electrical activity whose spectra overlap with that of uterine activity useful information encoded in the filtered frequency range is lost \cite{hassan2011combination,xu2022review}.

Ground electrodes and bipolar electrode arrangements remove electrical activity that is shared among multiple electrodes. For example, a ground electrode can be placed on the pregnant mothers' hip for common mode rejection, removing power line noise \cite{alexandersson2015icelandic}. Additionally, the measurements of one or more abdominal electrodes can be subtracted from the measurements of the rest of the abdominal electrodes to remove shared electrical activity \cite{hassan2011combination,jager2018characterization}. The differences between the measurements of these electrodes are known as bipolar measurements, and this procedure usually increases the signal-to-noise ratio (SNR) of the measured uterine activity \cite{hassan2011combination}. However, bipolar measurements have limitations \cite{hassan2011combination}. For example, bipolar measurements are significantly dependent on electrode configurations, including electrode positions, orientations, and interelectrode distances \cite{rooijakkers2014influence}.

Other methods have been developed for estimating uterine activity by denoising EHG measurements. For example, wavelet filtering and empirical mode decomposition (EMD) have been proposed for denoising EHG measurements \cite{leman2000rejection,limem2015uterine,taralunga2015empirical,huang1998empirical}. However, these methods do not fully overcome the limitations of spectral filtering approaches. Hassan et al.\ introduced a combination of canonical correlation analysis (CCA) and EMD to denoise EHG measurements \cite{hassan2011combination}. However, the components obtained through CCA and EMD do not have clear physiological interpretations, which complicates the selection of the various hyperparameters and design choices required for this method.

Tensor methods are compelling for manipulating multi-electrode EHG recordings. Whereas the methods mentioned above operate either on the measurements of each electrode independently or on a matrix arrangement of the measurements of the electrodes, tensor methods can operate directly on the measurement data, preserving the spatial and temporal structures of the measurements.
Statistical tensor decompositions separate complex multi-dimensional data into simpler components \cite{sidiropoulos2017tensor,cong2015tensor,cichocki2015tensor}.
By setting appropriate constraints, it is possible to regulate which properties of the measurements are captured in each of the basic components \cite{sidiropoulos2017tensor,cong2015tensor,cichocki2015tensor}. This approach is highly flexible, and a wide range of properties can be used to decompose the measurements, including spatial, temporal, and spectral properties \cite{cong2015tensor,cichocki2015tensor}. 
Notably, in applications using tensor measurements, statistical tensor decompositions have been shown to estimate the underlying data components more accurately than algorithms based on matrix unfolding, where tensor measurements are rearranged into a matrix form by stacking different dimensions along the rows or columns \cite{cichocki2015tensor}.

Previously, using simulated EHG measurements, Zahran et al.\ applied tensor methods to localize electrical sources within the uterus \cite{zahran2019performance}. Within physiological applications, tensor methods are most commonly used in processing electroencephalography (EEG) measurements, including extracting useful electrical activities \cite{cong2015tensor,tang2018bayesian}.

Here, we develop a statistical tensor decomposition method to estimate both electrical activity that is more variable through time and space, which we term localized activity, and electrical activity that is more uniformly distributed among the abdominal electrodes throughout the recording. Using simulated and real EHG measurements, we show that our method can estimate both localized and distributed activities more accurately than existing methods. Moreover, we show that the localized activity estimated with our method captures the uterine electrical activity during contractions with superior SNR than several filtering and denoising methods, while, together with the distributed activity, preserving all the valuable information in EHG recordings. Our method is useful for processing uterine electrical activity and, ultimately, for developing more effective applications to monitor uterine activity.

\section*{Methods}

\subsection*{EHG measurements used in this work}

We developed our work using EHG measurements from three sources. First, we simulated an EHG recording made with a four-by-four electrode grid placed over a pregnant mother's abdomen. These simulated measurements were made by adding various simple waveforms, representing physiological processes that can be observed in EHG measurements. These simulated measurements, detailed in the next section, were used to evaluate our method's performance against a ground truth. 

Second, we used unipolar EHG measurements from ``The Icelandic 16-electrode electrohysterogram database (Icelandic EHG database)," which includes EHG recordings from pregnant mothers during their third trimester of pregnancy (n=112) and during labor (n=10) \cite{alexandersson2015icelandic}. The recordings of mothers during labor in this dataset were defined by Alexandersson et al.\ as recordings made on mothers suspected to be in labor, who were in labor wards, and who gave birth within 24 hours \cite{alexandersson2015icelandic}. These measurements were made using a four-by-four electrode grid, where adjacent electrodes were placed 17.5 mm apart on the mothers' abdomens, and a ground and a reference electrode were placed on their iliac crests. Third, we used bipolar measurements from the ``Term-Preterm EHG dataset with tocogram (TPEHGT DS)," which includes 26 EHG measurements from pregnant mothers recorded around their 30th week of gestation \cite{jager2018characterization}. These recordings were made with a two-by-two electrode grid, where adjacent electrodes were 70 mm apart, and a ground electrode was placed on the mothers' thighs. In this dataset, the measurements were recorded as differences between abdominal electrodes.
We supplemented the three bipolar signals in these recordings with a fourth signal, as shown in \cite{jager2020assessing}. We stored these recordings as two-by-two-by-$T$ tensors, where $T$ is the temporal length of the recordings. The recordings in these two datasets were supplemented with clinical information, simultaneous tocograms, and annotations of when the mothers perceived uterine contractions. Whereas the TPEHGT DS contains annotations clearly delimiting the beginnings and endings of contractile intervals, containing EHG bursts, and non-contractile intervals (termed “dummy” intervals in the dataset), the annotations in the Icelandic EHG database do not delimit the beginnings and endings of contractile intervals.

\subsection*{Model assumptions and justification}

Our model is based on the rationale behind using bipolar EHG measurements to enhance the SNR of the recorded uterine activity. Bipolar measurements aim to remove the electrical components shared among the electrodes, for instance cardiac and respiratory activity, keeping only the electrical activity unique to each electrode \cite{hassan2011combination}. Following this logic and observing that the uterine electrical activity is highly non-stationary, we propose a sparse tensor to localize the electrical activity that is highly variable among electrodes and during the entire recording \cite{duchene1995analyzing}.

Besides localized activity, EHGs capture measurement noise and electrical activity that results from physiological processes. The measured physiological processes can be separated from random noise using low-rank structures \cite{sidiropoulos2017tensor}. Therefore, we use a low-rank tensor to estimate the electrical activity that is neither localized in time or space nor random noise in nature, which we refer to as distributed activity \cite{sidiropoulos2017tensor,nguyen2012denoising}.

\subsection*{Related work from video signal processing}

We make an analogy between the problem of estimating localized and distributed uterine electrical activity and the problem of separating a moving foreground from a static background in video signal processing.
In video signal processing, statistical tensor decomposition methods segment the foreground and background very well \cite{zhao2015bayesian,zhou2019bayesian,huang2021bayesian,huang2021bayesian}. Recent methods have alleviated some of the challenges of low-rank tensor decomposition, including its characteristically high computational complexity and the difficulty of estimating the low-rank model parameters together with the rank of the tensor \cite{zhou2019bayesian,zhao2015bayesian,zhao2015bayesian2}. For example, Zhao et al.\ developed a robust Bayesian tensor decomposition that separates the foreground from the background activity in videos with very good accuracy and computational speed \cite{zhao2015bayesian}. Such methods are compelling for the analysis of tensor EHG measurements. However, they are optimized for videos with mostly static backgrounds, while the distributed electrical activity in EHG measurements is non-stationary \cite{diab2021performance}.

Here, we represent EHG measurements as tensors. This tensor representation arises naturally in multi-electrode measurements, where the first two dimensions represent the spatial locations of the electrodes and the third dimension corresponds to the acquisition time. Based on this representation, we reformulate the Bayesian tensor decomposition method introduced by Zhao et al.\ for video signal processing (BRTF) to estimate localized and distributed electrical activity \cite{zhao2015bayesian}. Importantly, to model the background activity we use a Tucker tensor framework, which can be regarded as a more complex generalization of the CANDECOMP/PARAFAC (CP) framework used by Zhao et al. \cite{zhao2015bayesian,sidiropoulos2017tensor}.
We replaced the CP for the more flexible Tucker tensor framework because the distributed electrophysiological activity in EHG measurements is more variable over time than the backgrounds in the videos analyzed by Zhao et al.\ \cite{zhao2015bayesian}. Additionally, the Tucker tensor framework has been shown to outperform the CP framework in separating froward from background activity in videos \cite{huang2021bayesian}.

\subsection*{Model overview}

We represent the electric potentials obtained from a pregnant mother's abdomen using an $m\times n$ grid of electrodes during $T$ time-steps with the measurement tensor $\mathcal{Y}\in  \mathbb{R}^{m\times n \times T}$. We seek to decompose this measurement tensor into the sum of a sparse tensor of localized activity $(\mathcal{S}\in  \mathbb{R}^{m\times n \times T})$, a low-rank tensor of distributed activity $(\mathcal{X}\in  \mathbb{R}^{m\times n \times T})$, and a tensor of white Gaussian noise $(\mathcal{E}\in  \mathbb{R}^{m\times n \times T})$, i.e.,

\vspace{-2mm}

\begin{equation} \label{robust}
  \mathcal{Y} = \mathcal{S} + \mathcal{X} + \mathcal{E}. 
\end{equation}

The derivation and implementation of our method are detailed in the Supplementary methods. Briefly, we first represent the EHG measurements as a random process that results from a generative model that includes a sparse and a low-rank tensor. Then, we use a variational-Bayes approach to estimate the posterior distribution of the latent variables of the model. This estimation approach separates the original EHG measurement tensor $(\mathcal{Y}\in  \mathbb{R}^{m\times n \times T})$ into three additional tensors: a sparse tensor capturing localized activity $(\mathcal{S}\in  \mathbb{R}^{m\times n \times T})$, a low-rank tensor containing distributed activity $(\mathcal{X}\in  \mathbb{R}^{m\times n \times T})$, and a tensor of white Gaussian noise $(\mathcal{E}\in  \mathbb{R}^{m\times n \times T})$.

\section*{Results}

\subsection*{Tensor decomposition of simulated EHG measurements}

We first evaluated the performance of our algorithm using simulated EHG tensor measurements. We modeled an EHG recording, which includes localized and distributed electrical activity, to evaluate the performance of our method, as well as various alternative approaches, against a ground truth. This ground truth enables us to directly quantify the accuracy of various methods in estimating localized and distributed activities. This performance evaluation is complemented by the next section of our manuscript, where we evaluate the performance of our method using real EHG measurements.

\begin{figure}[!h]
    \centering
    \includegraphics[width =0.95\columnwidth]{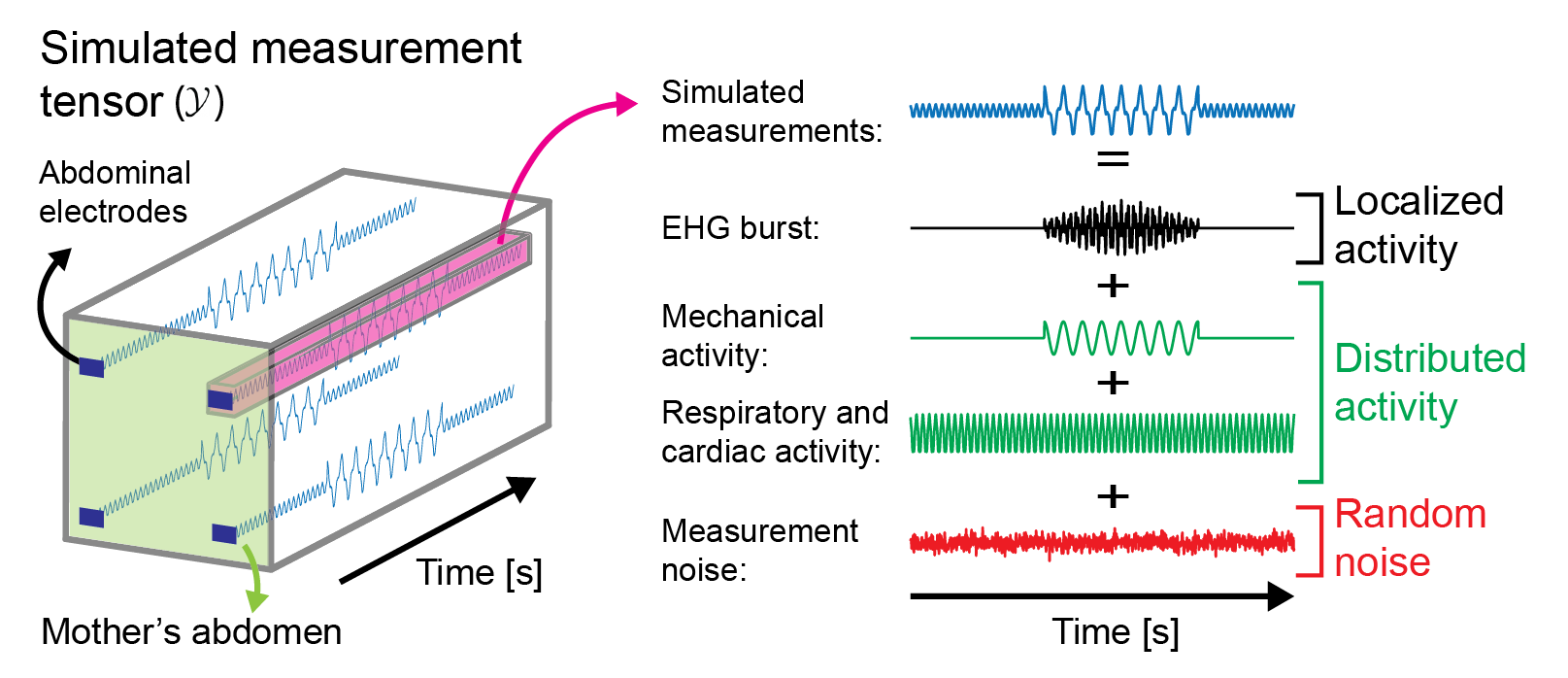}
    \caption{Simulated EHG tensor measurements. Each abdominal electrode records multiple physiological phenomena, which are illustrated on the right side of the figure. The waveforms shown on the right have dissimilar amplitude scale ranges.}
    \label{fig:simulated_description}
\end{figure}

We simulated a recording made with the electrode grid used by Alexandersson et al., that is, a four-by-four grid of monopolar electrodes, with an interelectrode distance of 17.5 mm, placed on a pregnant mother's abdomen \cite{alexandersson2015icelandic}. We simplified the components observed in EHG measurements using sine waves with characteristics that are consistent with reported values. We show a schematic representation of the simulated measurements in Fig.\ \ref{fig:simulated_description}.

The measurements in each electrode contain the sum of four components. The first measurement component recreates a travelling EHG burst and is the sum of two sine waves, representing the FWL and the FWH. These sine waves have frequencies of 0.2 and 0.6 Hz, and amplitudes of 0.2 and 0.3 mV, respectively \cite{batista2016multichannel,terrien2007spectral}. These frequencies align with previously reported values, and the amplitudes are consistent with both the energy ratio between the FWL and the FWH reported by Terrien et al.\ and previously reported amplitudes \cite{batista2016multichannel,terrien2007spectral,garfield2021review}. Additionally, this burst lasts for 80 seconds, which is in agreement with the mean duration of the annotated EHG bursts in the TPEHGT DS and the mean duration of contractions reported by Ye-Lin et al.\ \cite{jager2018characterization,ye2015feasibility}. Furthermore, the amplitude of this component is modulated by a Gaussian envelope, so that the peak amplitude of this component is 0.5 mV, and the amplitude gradually decreases towards the ends of the burst. 
This simulated waveform travels at 4.0 cm/s from the upper to the lower electrodes in the grid \cite{rabotti2010noninvasive,mikkelsen2013electrohysterography}. We simulated the waveform’s propagation by delaying the waveform across rows of electrodes, as shown by Rabotti et al.\ \cite{rabotti2010noninvasive}. Since this component varies over time and across electrodes, it is deemed to be localized activity.

The next two measurement components simulate distributed electrical activity. The second component simulates an SW, which has been shown to be related to uterine mechanical activity \cite{garfield2021review,devedeux1993uterine}. We modeled this component using a common sine wave across all electrodes, with a frequency of 0.02 Hz and an amplitude of 7.0 mV, during the time when the EHG burst is recorded by the electrode grid \cite{garfield2021review,batista2016multichannel}.

The third component simulates maternal respiratory and cardiac activity. We simulate the respiratory activity using a sine wave with a frequency of 0.3 Hz and a varying amplitude across the electrodes \cite{xu2022review,batista2016multichannel}. The respiratory component may have varying amplitudes in different electrodes depending on the electrodes’ placement and the mother's posture, among other factors. To account for possible amplitude variations throughout the electrode grid, we model this measurement component having a decreasing amplitude along the main diagonal. The respiratory activity has a maximal amplitude of 3.0 mV in the upper-left electrode and a minimal amplitude of 1.5 mV in the bottom-right electrode. Additionally, we simulated the maternal cardiac activity using a sine wave with a frequency of 1.2 Hz and an amplitude of 0.03 mV \cite{batista2016multichannel,xu2022review}. We set the amplitudes of the respiratory and cardiac activities based on the power ratios between these activities and uterine activity reported by Batista et al.\ \cite{batista2016multichannel}.

The fourth component recreates measurement noise and is simulated as white Gaussian noise. This noise has zero mean and variance such that the SNR of the measured signal is 15 dB, which is usual in electrophysiological recordings \cite{taralunga2015empirical}.

Lastly, the simulated measurements in each electrode consist of the sum of these components. In these measurements, the waveform created by the travelling EHG burst is concealed among the other measurement components. These measurements are stored as a tensor, where the first two dimensions encode the electrode position and the third dimension corresponds to the acquisition time.

\begin{figure}[!h]
    \centering
    \includegraphics[width =0.92\linewidth]{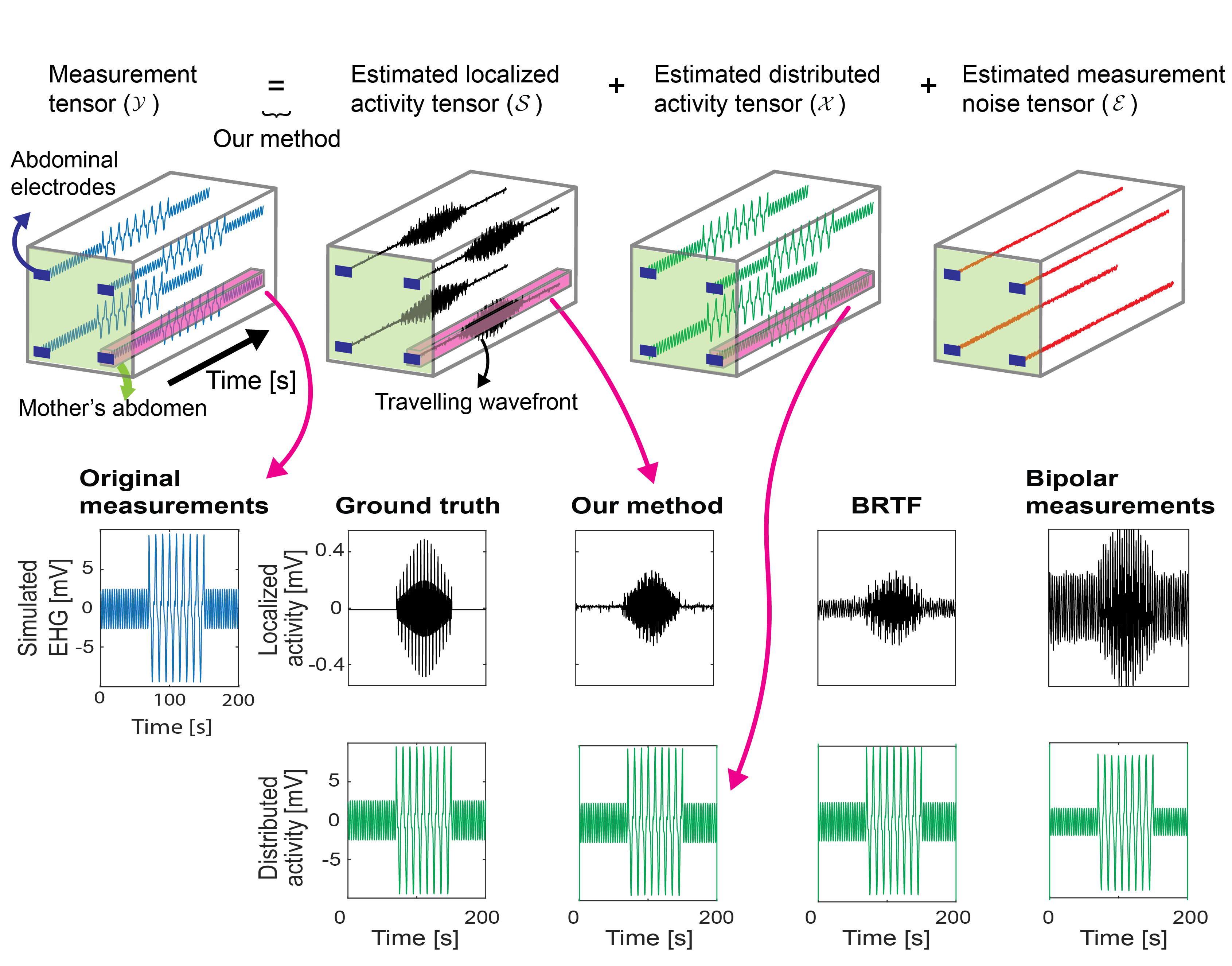}
    \caption{Tensor decomposition of the simulated EHG measurements. We estimated the localized and distributed electrical activities represented in Fig.\ \ref{fig:simulated_description}. The boxes in the upper part of the diagram show the results obtained with our method. The waveforms shown in the upper part have dissimilar amplitude scale ranges across the different tensors.
    The graphs on the bottom part of the diagram compare the localized and distributed activities estimated with our method, the second-best performing method (BRTF), and bipolar measurements. The measurements and the results obtained with our method for the 16 electrodes are shown in Supplementary Fig.\ 1.}
    \label{fig:estimated_description}
\end{figure}

\begin{figure}[!h]
    \centering
    \includegraphics[width =0.4\linewidth]{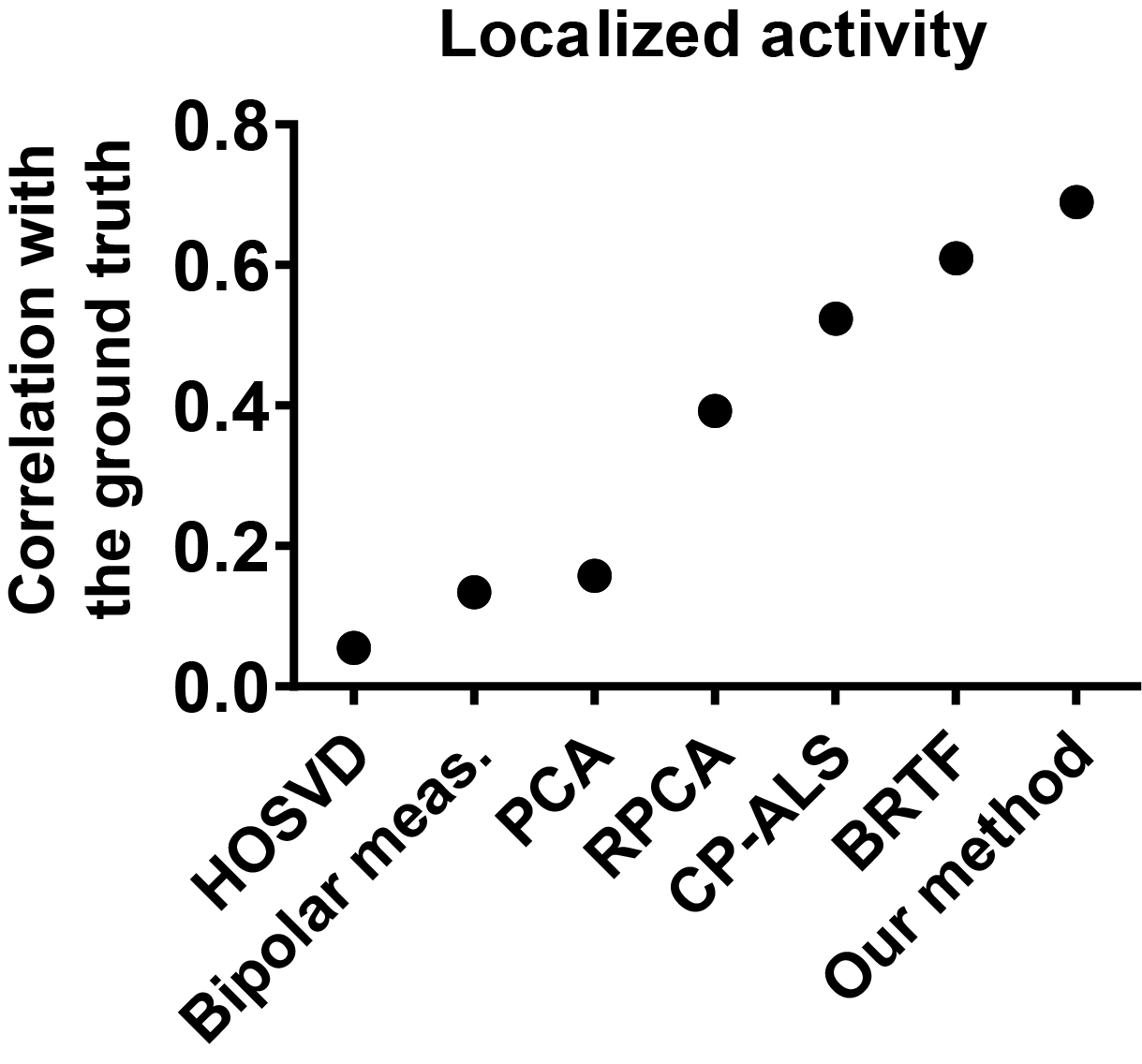}
    \caption{Pearson's correlation coefficients between the localized activity estimated by various methods and the corresponding ground truth. For the methods that have stochastic components, we present the average results of 100 runs. The sizes of the 95\% confidence intervals are smaller than the dots on the plot, and therefore, the intervals are not shown.}
    \label{fig:comparisson_of_methods}
\end{figure}

Using our method, we estimated the localized and distributed activities from the simulated EHG tensor measurements and we present the results in Figs.\ \ref{fig:estimated_description}, \ref{fig:comparisson_of_methods}, and Supplementary Fig.\ 1.
In the upper part of Fig.\ \ref{fig:estimated_description}, we illustrate how our method decomposes the measurement tensor into tensors of localized activity, distributed activity, and measurement noise. Notably, although the localized activity can be identified using bipolar measurements, BRTF and our method estimate the localized activity more accurately, as shown in the lower part of Fig.\ \ref{fig:estimated_description}.

We decomposed the simulated EHG measurements described in the previous subsection using multiple methods. Our method recovered the localized and distributed activities more accurately than these methods, as shown in the lower part of Fig.\ \ref{fig:estimated_description} and in Fig.\ \ref{fig:comparisson_of_methods}. We first considered bipolar measurements, where the original measurements represent the distributed activity and the difference between adjacent rows of electrodes represent the localized activity.

Secondly, we estimated the localized and distributed activities using principal component analysis (PCA), which we used to compute a low-rank approximation of the measurements. Thirdly, we decomposed the simulated EHG measurements using two methods that approximate tensors using low-rank structures (CP-ALS and HOSVD) \cite{sidiropoulos2017tensor}. When using PCA, CP-ALS, and HOSVD, we considered the low-rank approximations as the distributed activity and the difference between the original measurements and the distributed activity as the localized activity. Lastly, we decomposed the simulated EHG using two approaches (RPCA and BRTF) that decompose measurement tensors into a low-rank tensor and a sparse tensor \cite{zhao2015bayesian,aravkin2014variational}. According to our assumptions, we consider the low-rank tensor to be the distributed activity and the sparse tensor to be the localized activity.

We compared the performance of these methods using the Pearson's correlation coefficient of the localized and distributed activities obtained with their respective ground truth measurements. Whereas all these methods were able to estimate the distributed activity with a correlation coefficient higher than 0.99 with the ground truth, these methods estimated the localized activity with varying performances, as shown in Fig.\ \ref{fig:comparisson_of_methods}. For each method besides the bipolar measurements, we fine-tuned the model's hyperparameters to obtain the best possible results.

\clearpage

\subsection*{Tensor decomposition of real EHG measurements}

\begin{figure}[!h]
    \centering
    \includegraphics[width =0.9\linewidth]{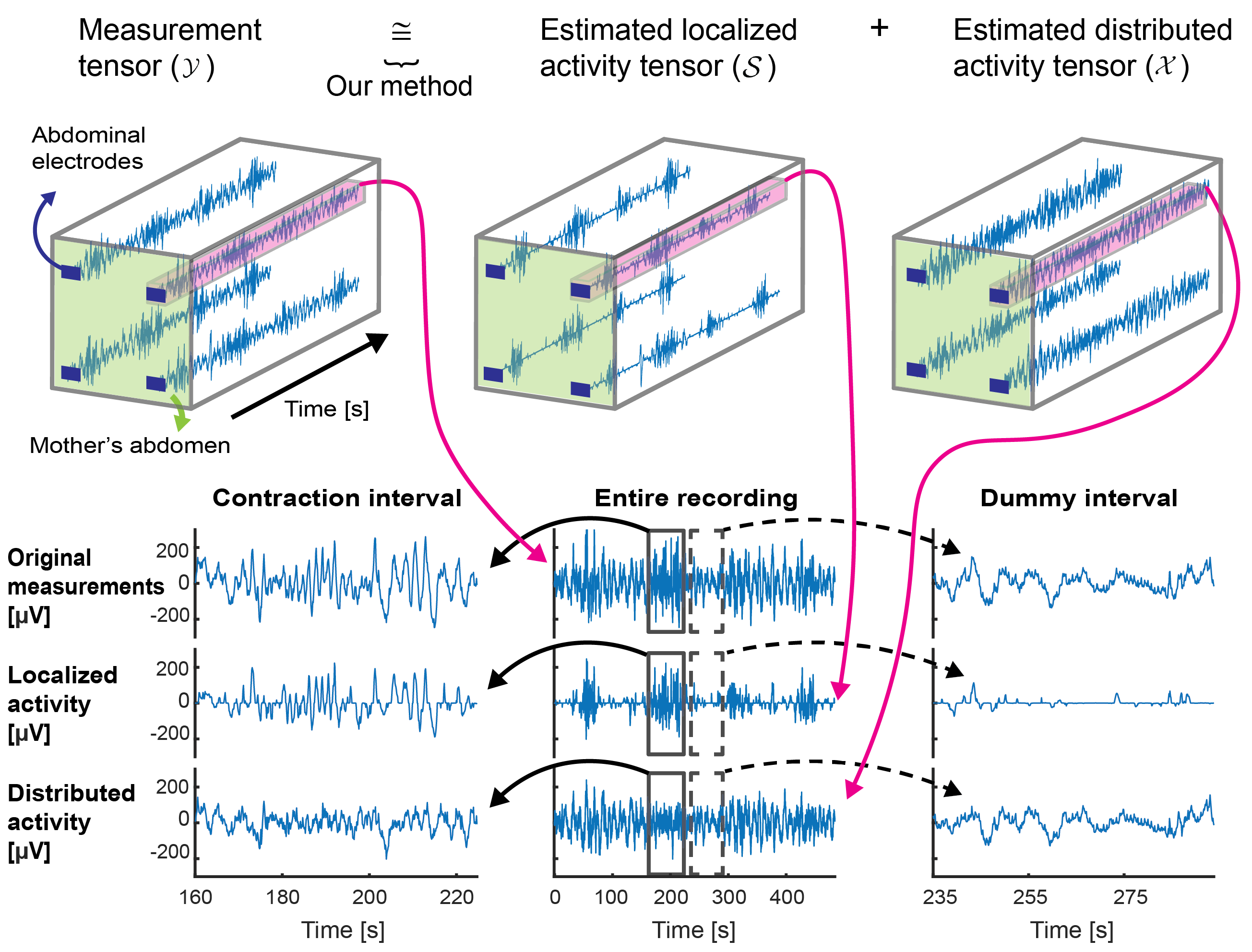}
    \caption{Localized and distributed electrical activity estimation from an EHG recording from the Icelandic EHG database. The localized activity estimated with our method captures the uterine contractile activity, while the distributed activity captures the electrical activity shared among the electrodes. The contraction and dummy intervals shown here were identified using the accompanying tocodynamometer recording, which is provided as Supplementary Fig.\ 2. The measurements from all 16 electrodes are shown in Supplementary Fig.\ 3.}
    \label{fig:real_decomposition}
\end{figure}

\begin{figure}[!h]
    \centering
    \includegraphics[width =0.9\linewidth]{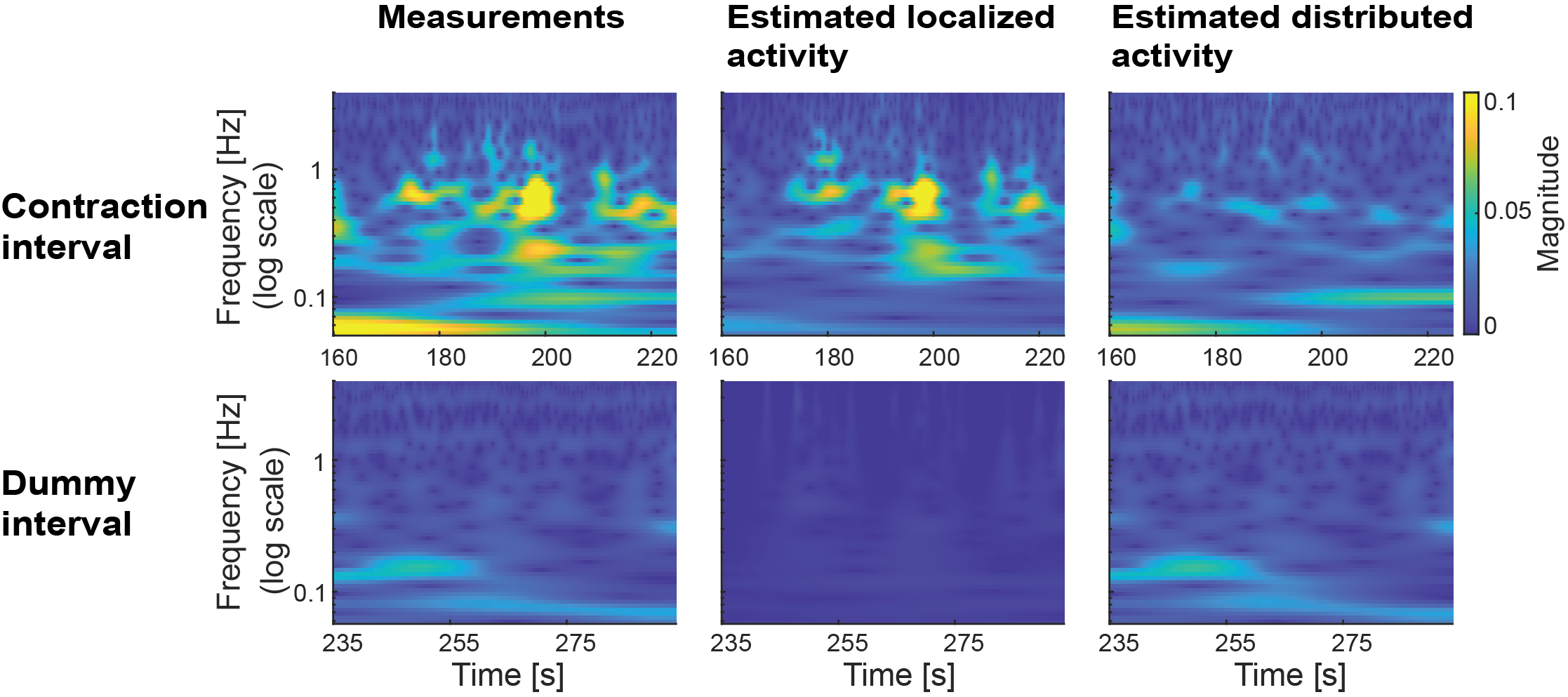}
    \caption{Scalograms of the measurements shown in Fig.\ \ref{fig:real_decomposition}.   
    The panels in the top row, from left to right show the scalograms of the original measurements, the estimated localized activity, and the estimated distributed activity for the contraction interval shown in Fig.\ \ref{fig:real_decomposition}. These panels share the same temporal ranges.
    The panels in the bottom row show the corresponding scalograms for the dummy interval shown in Fig.\ \ref{fig:real_decomposition}. The bottom panels share the same temporal ranges. All panels share the same magnitude and logarithmic frequency scales.}
    \label{fig:spectrum}
\end{figure}

In Figs.\ \ref{fig:real_decomposition}, \ref{fig:spectrum}, and Supplementary Fig.\ 3, we show the localized and distributed electrical activities estimated with our method from a labor EHG recording in the Icelandic EHG database. Before estimating the localized and distributed activities, we first removed the initial minute of the measurements, eliminating transient effects at the beginning of the recording. Next, we filtered the original measurements using a bidirectional bandpass Butterworth filter with cutoff frequencies of 0.05 Hz and 4.0 Hz to remove baseline wander and high frequency electrical activity that is not associated with uterine activity \cite{jager2018characterization}. Afterwards, we downsampled the recordings to 10 Hz to improve the computational speed without losing information. We filtered and downsampled the measurements along the temporal dimension, i.e., we preprocessed the measurements from each electrode independently.

The EHG recording shown in Fig.\ \ref{fig:real_decomposition} contains four uterine contractions, which can be clearly identified as four separate bursts in the tensor of the estimated localized activity for the electrodes. In contrast, these bursts of electrical activity cannot be clearly distinguished in the tensor of original measurements. In the bottom part of Fig.\ \ref{fig:real_decomposition}, we observe that the localized activity estimated with our method preserves the waveforms associated with uterine contractions, while dampening the electrical activity measured between contractions, i.e., during dummy intervals. The localized activity captures the signal deflections which are unique to each electrode and are not part of a stationary process throughout the duration of the recording. Moreover, the distributed activity captures the rest of the physiological activity, which is more uniform throughout the recording.

In Fig.\ \ref{fig:spectrum}, we present the scalograms of the measurements shown in Fig.\ \ref{fig:real_decomposition}. Consistent with the results shown in Fig.\ \ref{fig:real_decomposition}, most of the energy during the contractile interval is captured by the localized activity, while most of the energy during the dummy interval is contained in the distributed activity. The scalogram of localized activity during the contraction interval shows activity which is variable over time. Conversely, the scalograms of distributed activity are less variable over time and present similar patterns during both the contraction and dummy intervals. The scalograms of original measurements and localized activity during the contraction interval resemble the scalograms of contractile activity reported by Hassan et al.\ \cite{hassan2011combination}.

\begin{figure}[!h]
    \centering
    \includegraphics[width =\linewidth]{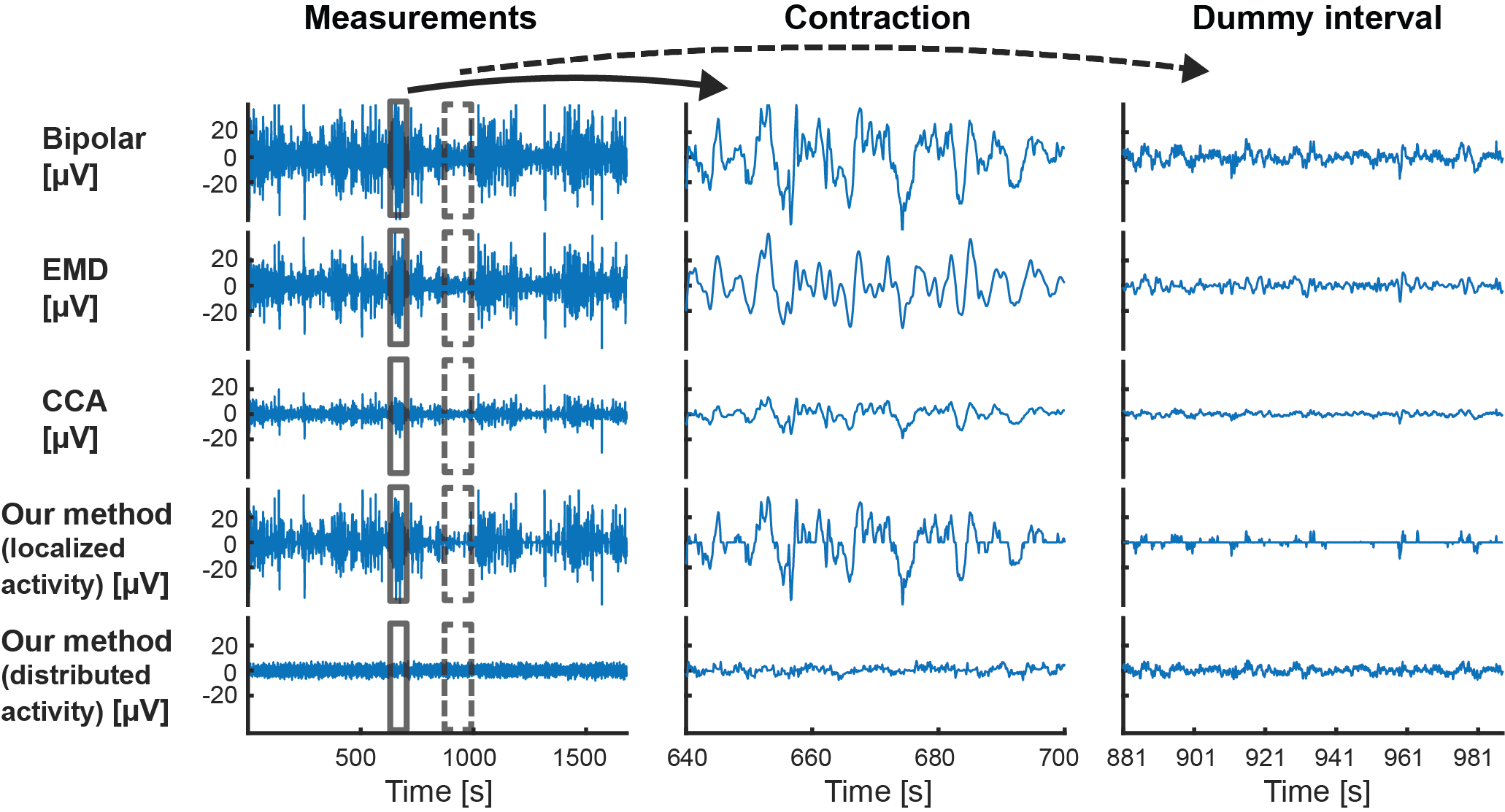}
    \caption{Comparison of uterine electrical activity estimation using bipolar measurements, EMD, CCA, and our method. The plots on the left column show the entire measurements from the same recording in the TPEHGT DS, but processed with different methods. The upper left plot shows the original bipolar recording. The middle and right columns show magnified views of a contraction and dummy interval annotated in the dataset, respectively. All the subpanels share the same amplitude scale range.} 
    \label{fig:real_comparisson}
\end{figure}

In Fig.\ \ref{fig:real_comparisson}, we compare our method with three methods reported for identifying uterine activity, namely bipolar measurements, EMD, and CCA, as described in \cite{hassan2011combination}. Here, for comparison, we magnify a contractile and a dummy interval from a bipolar EHG recording from the TPEHGT DS, which was first preprocessed as described before, and present it along with the measurements obtained by applying our method, EMD, and CCA. For all these methods, we fine-tuned their hyperparameters to obtain the best possible results.

Similar to our observations from Fig.\ \ref{fig:real_decomposition}, in Fig.\ \ref{fig:real_comparisson}, we observe that the localized activity estimated with our method successfully preserves the prominent deflections measured during a uterine contractile interval, while dampening most of the electrical activity measured during a dummy interval. Given that we estimate the localized activity using a sparse tensor and the distributed activity with a low-rank tensor, the localized activity captures the deflections that are unique to each electrode and cannot be modeled accurately with a low-rank structure, while zeroing the activity that is better captured as distributed activity. Moreover, the distributed activity captures more uniform electrical activity measured throughout the recording, including during contraction intervals. The distributed activity in this example has a lower amplitude than the localized activity because these measurements were obtained with bipolar electrodes, which partially filter the distributed activity.

In contrast, the contractile and dummy waveforms obtained with EMD contain only the low frequency components observed in the bipolar recordings, as explained in \cite{hassan2011combination}. Lastly, although the contractile activity recorded with bipolar electrodes has a considerably higher amplitude than the electrical activity recorded during a dummy interval, after applying CCA the electrical activities measured both during the contraction and dummy intervals have more similar amplitudes.

\begin{figure}[!h]
    \centering
    \includegraphics[width =0.8\columnwidth]{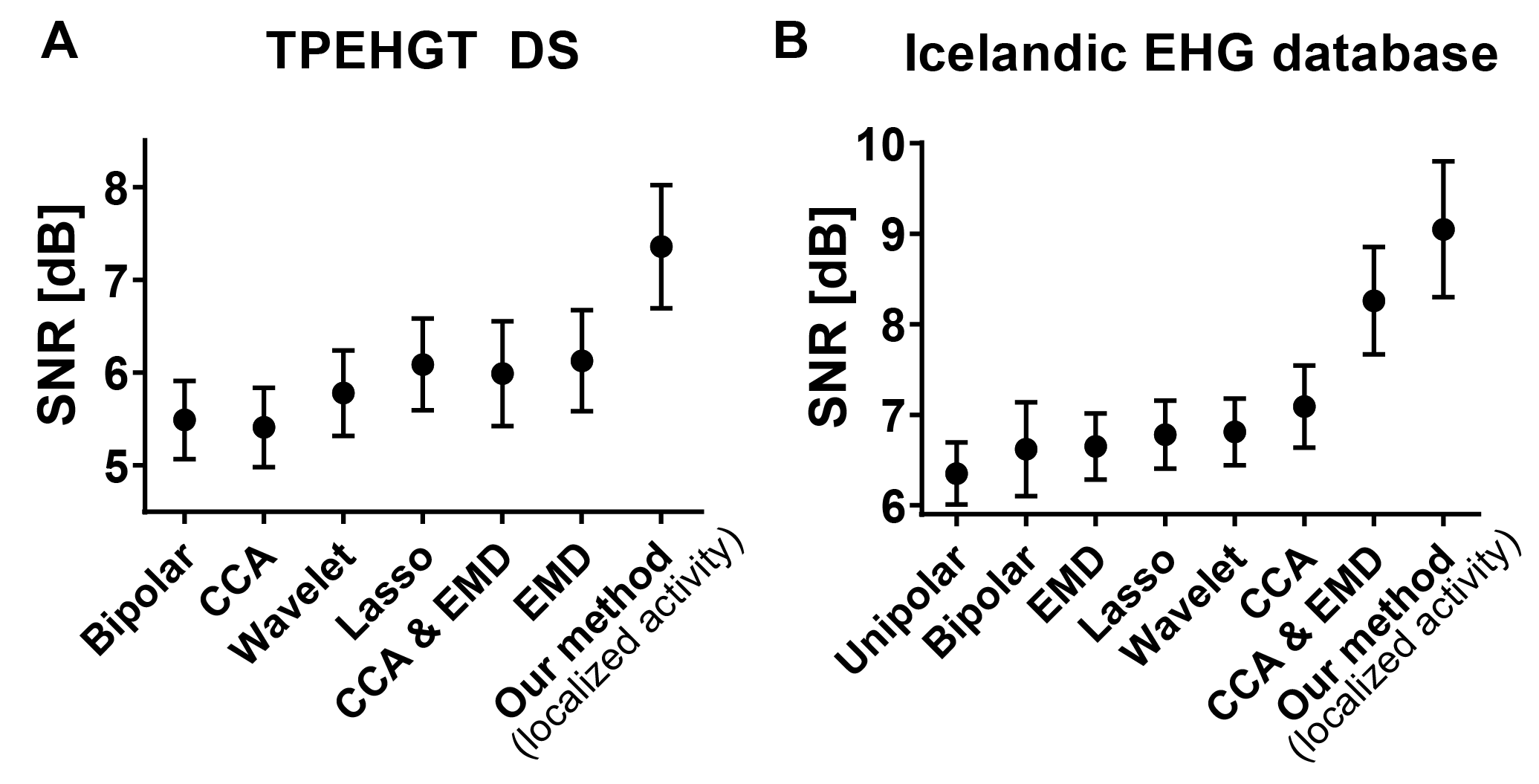}
    \caption{SNRs of uterine activity obtained with various methods. A. Average SNRs obtained across all electrodes and recordings in the TPEHGT DS for various methods. Error bars indicate 95\% confidence intervals. B. Similar to A, but using recordings from the Icelandic EHG database.}
    \label{fig:SNR_comparisson}
\end{figure}

In Fig.\ \ref{fig:SNR_comparisson}, we quantify the SNR obtained for two datasets with our method and with various alternative methods for denoising EHG measurements. Besides our method, we denoised the measurements using EMD and CCA, separately and in combination, as described in \cite{hassan2011combination}, wavelet filtering, as described in \cite{limem2015uterine}, and Lasso denoising, as described in \cite{chambolle2010introduction,L1_matlab}. We calculated the SNR as the power ratio between uterine contraction intervals and dummy intervals over all the electrodes, using a similar approach as Hassan et al.\ \cite{hassan2011combination}. When using measurements from the TPEHGT DS, we considered the contractile and dummy intervals provided in the annotations. Since the recordings in this dataset are provided as bipolar measurements, in this case the denoising methods were applied on bipolar measurements. When using measurements from the Icelandic EHG database, we identified contractile and dummy intervals based on visual inspection aided by the simultaneous tocograms. To facilitate the identification of contraction and dummy intervals and to avoid calculating SNRs using mislabeled contraction or dummy intervals, we considered only eight EHG measurements in the Icelandic EHG database recorded during labor, which present stronger tocogram and EHG bursts during contractions.

In Fig.\ \ref{fig:SNR_comparisson}, we observe that the localized activity obtained with our method captures uterine contractile activity with a higher SNR than other methods commonly used for this task. Moreover, we observe that while most of the methods usually used for denoising EHG and other physiological measurements achieve only moderate signal quality improvements, our method significantly increases the SNR of uterine activity.

\section*{Discussion}

We developed a framework for analyzing multielectrode EHG measurements. Specifically, we developed a Bayesian tensor decomposition method for estimating localized and distributed electrical activity. Following the rationale for bipolar measurements and observing that uterine electrical activity has more power concentrated during contractile intervals, we used a sparse tensor to capture the electrical activity localized to each electrode and more variable during the recordings. Moreover, we employed a low-rank Tucker tensor to estimate the more stationary distributed electrical activity. Then, we evaluated the performance of our method using simulated and real EHG measurements. Finally, we showed that our method can estimate uterine activity in EHGs more accurately than other methods.

Our experiments with simulated EHG measurements show the advantage of tensor methods in general, and of our method in particular, in estimating localized and distributed electrical activities. Although we simulated EHG measurements using simple waveforms, our simulated data recreates many measurement components present in EHG measurements and reproduces the challenge of estimating localized and distributed electrical activities. These simulated measurements are useful because their simulation generates ground truth data that can be used for performance evaluation. Using simulated data, we illustrated the limitations of bipolar measurements and showed that, while various methods can estimate the distributed electrical activity accurately, estimating the localized activity is more challenging. Moreover, we showed how our method can estimate this localized activity more accurately than other methods. We hypothesize that our method outperformed BRTF in this task because the Tucker tensor framework is more adaptable than the CP decomposition to the non-stationary EHG measurements \cite{zhao2015bayesian2,sidiropoulos2017tensor}. This is supported by the fact that the Tucker tensor framework has been shown to outperform the CP decomposition in video signal processing applications \cite{huang2021bayesian}.

Using real EHG measurements, we showed that our method can separate uterine activity from distributed electrical activity more accurately than other methods commonly used for this task. In Figs.\ \ref{fig:real_decomposition}, and \ref{fig:spectrum}, and \ref{fig:real_comparisson}, we illustrate how the localized electrical activity estimated using our method captures the uterine contractile activity observed in unipolar and bipolar measurements, while dampening the additional electrical activity from other sources. Additionally, we show that the distributed electrical activity estimated with our method captures the shared physiological electrical activity that is present throughout the recording, removing measurement noise.

Furthermore, in Fig.\ \ref{fig:SNR_comparisson} we quantify the advantage of our method for estimating uterine electrical activity during contractions by using measurements from two different datasets. We also show that our method outperforms various other methods commonly used for this task. 
Our comparison includes the combination of CCA with EMD, which has been reported to produce satisfactory results \cite{hassan2011combination}. Our implementation of CCA and EMD achieves comparable SNR values on unipolar measurements to those reported previously, although the original unipolar measurements used in our work have a higher SNR than the unipolar measurements used by Hassan et al.\ \cite{hassan2011combination}. Notably, the combination of CCA and EMD for denoising unipolar measurements outperforms each of these methods separately, suggesting that our method could also be combined with other methods for further denoising. Moreover, our method is able to estimate uterine activity accurately from both unipolar and bipolar measurements, suggesting that it could also be useful with other measurement configurations. 

Our method does not depend on the amplitude or the frequency of uterine electrical activity. Accordingly, our method performed well on our simulated data and on two datasets containing contractions with different characteristics. In our simulated data some components of the distributed activity have higher amplitudes than the EHG burst. Additionally, the datasets we used capture contractions with different characteristics. As shown in Fig. \ref{fig:SNR_comparisson}, the contractions in the recordings we used from the Icelandic dataset have a higher SNR than the contractions in the TPEHGT DS. Furthermore, the contractions in both datasets have varying spectra. In the TPEHGT DS, the measurements have different spectra depending on whether they were recorded from mothers who delivered on term or preterm \cite{jager2018characterization}. Moreover, we used the measurements from the Icelandic database that were recorded during labor and therefore, have different spectra from the EHG recordings in the TPEHGT DS, which were measured during pregnancy \cite{jager2018characterization,terrien2010synchronization}.

Our method is not restricted to EHG measurements: it can be extended to a range of multidimensional data analysis applications. Importantly, we note that our method is based only on assumptions regarding the structure of the localized and distributed activities. However, our method does not rely on specific properties of EHG measurements, such as tensor dimensions, interactions between the distributed and localized activity, or the true multilinear rank of the distributed activity. Therefore, we expect that our method will be useful for a range of multidimensional data analysis applications, including analyzing electrophysiological data such as EEG recordings or multidimensional RNA sequencing data \cite{cong2015tensor,hore2016tensor}. Additionally, our method is applicable to a wider scope of multidimensional data applications, such as hyper-spectral imaging and video signal processing \cite{zhao2015bayesian,zhang2019computational}.

Our method has two main limitations with respect to other methods for estimating uterine activity. Although variational Bayes inference is an efficient approach for estimating the posterior distributions of the model parameters, estimating uterine activity with our method is more computationally complex than simpler methods, such as using bipolar measurements. Therefore, our method may not be appropriate if real-time measurements are needed. Additionally, our method is designed for abdominal electrodes that are arranged in a square lattice, and it may not be directly applicable for more complicated arrangements of electrodes.

In the future, rather than processing the entire recording in one batch, we plan to expand our method to estimate localized and distributed activities adaptively during acquisition, enabling real time monitoring. Additionally, our method may be expanded for processing data stored in graphs, rather than tensors, to process EHG measurements recorded with more complex electrode arrangements. Furthermore, our method could be further validated using more complex simulations of EHG measurements and using simultaneously recorded invasive measurements of myometrial electrical activity and EHGs \cite{laforet2011toward,wu2019noninvasive,la2012multiscale,zhang2016modeling}.

Finally, our method could be useful for investigating uterine electrophysiology and for developing applications to inform obstetric care. For example, our method could be used for characterizing different types of uterine contractions and varying distributed electrical activities \cite{esgalhado2020uterine}. Additionally, our method can assist in denoising EHG measurements, extracting informative features, and revealing components of EHG measurements to develop practical applications, such as predicting preterm births and detecting imminent labors \cite{xu2022review}.

\section*{Conclusion}

We developed a tensor framework for estimating localized and distributed electrical activity in noninvasive EHG measurements. Our method addresses the inherent limitations of noninvasive electrophysiological measurements and enables the estimation of uterine activity more accurately than commonly used methods. By separating the physiological activity measured in EHG recordings into localized and distributed activity, our work can help in developing more effective applications based on EHG measurements.

\section*{Acknowledgement}

This research was supported by the McDonnell International Scholars Academy at Washington University in St.\ Louis.

\bibliographystyle{unsrtnat}

\bibliography{references}

\begin{thebibliography}{58}
\providecommand{\natexlab}[1]{#1}
\providecommand{\url}[1]{\texttt{#1}}
\expandafter\ifx\csname urlstyle\endcsname\relax
  \providecommand{\doi}[1]{doi: #1}\else
  \providecommand{\doi}{doi: \begingroup \urlstyle{rm}\Url}\fi

\bibitem[Hillemeier et~al.(2007)Hillemeier, Weisman, Chase, and
  Dyer]{hillemeier2007individual}
Marianne~M Hillemeier, Carol~S Weisman, Gary~A Chase, and Anne-Marie Dyer.
\newblock Individual and community predictors of preterm birth and low
  birthweight along the rural-urban continuum in central pennsylvania.
\newblock \emph{J. Rural Health}, 23\penalty0 (1):\penalty0 42--48, 2007.

\bibitem[Elixhauser and Wier(2011)]{elixhauser2011complicating}
Anne Elixhauser and Lauren~M Wier.
\newblock Complicating conditions of pregnancy and childbirth, 2008:
  Statistical brief \#113.
\newblock \url{https://www.ncbi.nlm.nih.gov/books/NBK56037/}, 2011.
\newblock Accessed: 2023-2-24.

\bibitem[Donna L.~Hoyert(2022)]{cdc_maternal_mortality}
National Center for Health Statistics~(U.S.) Donna L.~Hoyert.
\newblock Maternal mortality rates in the united states, 2020.
\newblock \url{https://stacks.cdc.gov/view/cdc/113967}, 2022.
\newblock Accessed: 2022-3-21.

\bibitem[{Centers for Disease Control and Prevention and
  others}(2017)]{centers2017severe}
{Centers for Disease Control and Prevention and others}.
\newblock Severe maternal morbidity in the united states.
\newblock
  \url{https://www.cdc.gov/reproductivehealth/maternalinfanthealth/severematernalmorbidity.html},
  2017.
\newblock Accessed: 2023-2-19.

\bibitem[Schneider et~al.(2014)Schneider, Group, et~al.]{schneider2014s1}
KTM Schneider, Maternal Fetal Medicine~Study Group, et~al.
\newblock S1-guideline on the use of ctg during pregnancy and labor.
\newblock \emph{Geburtshilfe und Frauenheilkunde}, 74\penalty0 (08):\penalty0
  721--732, 2014.

\bibitem[Gee et~al.(2020)Gee, Ma'ayeh, Ward, Buhimschi, Klebanoff, and
  Rood]{gee2020intrauterine}
Stephen~E Gee, Marwan Ma'ayeh, Calvin Ward, Catalin Buhimschi, Mark Klebanoff,
  and Kara Rood.
\newblock Intrauterine pressure catheter use is associated with an increased
  risk of postcesarean surgical site infections.
\newblock \emph{Am. J. Perinatol.}, 37\penalty0 (06):\penalty0 557--561, 2020.

\bibitem[Harper et~al.(2013)Harper, Shanks, Tuuli, Roehl, and
  Cahill]{harper2013risks}
Lorie~M Harper, Anthony~L Shanks, Methodius~G Tuuli, Kimberly~A Roehl, and
  Alison~G Cahill.
\newblock The risks and benefits of internal monitors in laboring patients.
\newblock \emph{Am. J. Obstet. Gynecol.}, 209\penalty0 (1):\penalty0 38--e1,
  2013.

\bibitem[Rood(2012)]{rood2012complications}
Kara~M Rood.
\newblock Complications associated with insertion of intrauterine pressure
  catheters: an unusual case of uterine hypertonicity and uterine perforation
  resulting in fetal distress after insertion of an intrauterine pressure
  catheter.
\newblock \emph{Case Rep., Obstet. Gynecol.}, 2012, 2012.

\bibitem[Eswaran et~al.(2004)Eswaran, Preissl, Wilson, Murphy, and
  Lowery]{eswaran2004prediction}
Hari Eswaran, Hubert Preissl, James~D Wilson, Pam Murphy, and Curtis~L Lowery.
\newblock Prediction of labor in term and preterm pregnancies using
  non-invasive magnetomyographic recordings of uterine contractions.
\newblock \emph{Am. J. Obstet. Gynecol.}, 190\penalty0 (6):\penalty0
  1598--1602, 2004.

\bibitem[Govindan et~al.(2015)Govindan, Siegel, Mckelvey, Murphy, Lowery, and
  Eswaran]{govindan2015tracking}
Rathinaswamy~B Govindan, Eric Siegel, Samantha Mckelvey, Pam Murphy, Curtis~L
  Lowery, and Hari Eswaran.
\newblock Tracking the changes in synchrony of the electrophysiological
  activity as the uterus approaches labor using magnetomyographic technique.
\newblock \emph{Reprod. Sci.}, 22\penalty0 (5):\penalty0 595--601, 2015.

\bibitem[Eswaran et~al.(2002)Eswaran, Preissl, Wilson, Murphy, Robinson, and
  Lowery]{eswaran2002first}
Hari Eswaran, Hubert Preissl, James~D Wilson, Pam Murphy, Stephen~E Robinson,
  and Curtis~L Lowery.
\newblock First magnetomyographic recordings of uterine activity with
  spatial-temporal information with a 151-channel sensor array.
\newblock \emph{Am. J. Obstet. Gynecol.}, 187\penalty0 (1):\penalty0 145--151,
  2002.

\bibitem[Wu et~al.(2019)Wu, Wang, Zhao, Talcott, Lai, McKinstry, Woodard,
  Macones, Schwartz, Cahill, et~al.]{wu2019noninvasive}
Wenjie Wu, Hui Wang, Peinan Zhao, Michael Talcott, Shengsheng Lai, Robert~C
  McKinstry, Pamela~K Woodard, George~A Macones, Alan~L Schwartz, Alison~G
  Cahill, et~al.
\newblock Noninvasive high-resolution electromyometrial imaging of uterine
  contractions in a translational sheep model.
\newblock \emph{Sci. Transl. Med.}, 11\penalty0 (483), 2019.

\bibitem[Xu et~al.(2022)Xu, Chen, Lou, Shen, and Pumir]{xu2022review}
Jinshan Xu, Zhenqin Chen, Hangxiao Lou, Guojiang Shen, and Alain Pumir.
\newblock Review on ehg signal analysis and its application in preterm
  diagnosis.
\newblock \emph{Biomed. Signal Process. Control}, 71:\penalty0 103231, 2022.

\bibitem[Rabotti et~al.(2008)Rabotti, Mischi, van Laar, Oei, and
  Bergmans]{rabotti2008estimation}
Chiara Rabotti, Massimo Mischi, Judith~OEH van Laar, Guid~S Oei, and Jan~WM
  Bergmans.
\newblock Estimation of internal uterine pressure by joint amplitude and
  frequency analysis of electrohysterographic signals.
\newblock \emph{Physiol. Meas.}, 29\penalty0 (7):\penalty0 829, 2008.

\bibitem[Rooijakkers et~al.(2014{\natexlab{a}})Rooijakkers, Rabotti, Oei,
  Aarts, and Mischi]{rooijakkers2014low}
Michael~J Rooijakkers, Chiara Rabotti, S~Guid Oei, Ronald~M Aarts, and Massimo
  Mischi.
\newblock Low-complexity intrauterine pressure estimation using the teager
  energy operator on electrohysterographic recordings.
\newblock \emph{Physiol. Meas.}, 35\penalty0 (7):\penalty0 1215,
  2014{\natexlab{a}}.

\bibitem[Mischi et~al.(2017)Mischi, Chen, Ignatenko, de~Lau, Ding, Oei, and
  Rabotti]{mischi2017dedicated}
Massimo Mischi, Chuan Chen, Tanya Ignatenko, Hinke de~Lau, Beijing Ding,
  SG~Guid Oei, and Chiara Rabotti.
\newblock Dedicated entropy measures for early assessment of pregnancy
  progression from single-channel electrohysterography.
\newblock \emph{IEEE. Trans. Biomed. Eng.}, 65\penalty0 (4):\penalty0 875--884,
  2017.

\bibitem[Hassan et~al.(2012)Hassan, Terrien, Muszynski, Alexandersson, Marque,
  and Karlsson]{hassan2012better}
Malunoud Hassan, Jeremy Terrien, Charles Muszynski, Asgeir Alexandersson,
  Catherine Marque, and Brynjar Karlsson.
\newblock Better pregnancy monitoring using nonlinear correlation analysis of
  external uterine electromyography.
\newblock \emph{IEEE. Trans. Biomed. Eng.}, 60\penalty0 (4):\penalty0
  1160--1166, 2012.

\bibitem[Goldsztejn and Nehorai(2022)]{goldsztejn2022predicting}
Uri Goldsztejn and Arye Nehorai.
\newblock Predicting preterm births from electrohysterogram recordings via deep
  learning.
\newblock \emph{medRxiv preprint}, pages 2022--12, 2022.
\newblock \doi{https://doi.org/10.1101/2022.12.25.22283937}.

\bibitem[Hassan et~al.(2011)Hassan, Boudaoud, Terrien, Karlsson, and
  Marque]{hassan2011combination}
Mahmoud Hassan, Sofiane Boudaoud, Jeremy Terrien, Brynjar Karlsson, and
  Catherine Marque.
\newblock Combination of canonical correlation analysis and empirical mode
  decomposition applied to denoising the labor electrohysterogram.
\newblock \emph{IEEE. Trans. Biomed. Eng.}, 58\penalty0 (9):\penalty0
  2441--2447, 2011.

\bibitem[Batista et~al.(2016)Batista, Najdi, Godinho, Martins, Serrano,
  Ortigueira, and Rato]{batista2016multichannel}
Arnaldo~G Batista, Shirin Najdi, Daniela~M Godinho, Catarina Martins,
  F{\'a}tima~C Serrano, Manuel~D Ortigueira, and Raul~T Rato.
\newblock A multichannel time--frequency and multi-wavelet toolbox for uterine
  electromyography processing and visualisation.
\newblock \emph{Comput. Biol. Med.}, 76:\penalty0 178--191, 2016.

\bibitem[Terrien et~al.(2007)Terrien, Marque, and
  Karlsson]{terrien2007spectral}
Jeremy Terrien, Catherine Marque, and Brynjar Karlsson.
\newblock Spectral characterization of human ehg frequency components based on
  the extraction and reconstruction of the ridges in the scalogram.
\newblock In \emph{2007 29th Conf. Proc. IEEE Eng. Med. Biol. Soc.}, pages
  1872--1875. IEEE, 2007.

\bibitem[Garfield et~al.(2021)Garfield, Murphy, Gray, and
  Towe]{garfield2021review}
RE~Garfield, Lauren Murphy, Kendra Gray, and Bruce Towe.
\newblock Review and study of uterine bioelectrical waveforms and vector
  analysis to identify electrical and mechanosensitive transduction control
  mechanisms during labor in pregnant patients.
\newblock \emph{Reprod. Sci.}, 28:\penalty0 838--856, 2021.

\bibitem[Ye-Lin et~al.(2014)Ye-Lin, Garcia-Casado, Prats-Boluda,
  Alberola-Rubio, and Perales]{ye2014automatic}
Yiyao Ye-Lin, Javier Garcia-Casado, Gema Prats-Boluda, Jos{\'e} Alberola-Rubio,
  and Alfredo Perales.
\newblock Automatic identification of motion artifacts in ehg recording for
  robust analysis of uterine contractions.
\newblock \emph{Comput. Math. Methods Med.}, 2014, 2014.

\bibitem[AYD{\i}N(2009)]{aydin2009comparison}
SERAP AYD{\i}N.
\newblock Comparison of power spectrum predictors in computing coherence
  functions for intracortical eeg signals.
\newblock \emph{Ann. Biomed. Eng.}, 37\penalty0 (1):\penalty0 192--200, 2009.

\bibitem[Rooijakkers et~al.(2014{\natexlab{b}})Rooijakkers, Song, Rabotti, Oei,
  Bergmans, Cantatore, and Mischi]{rooijakkers2014influence}
Michael~Johannes Rooijakkers, Shuang Song, Chiara Rabotti, S~Guid Oei, Jan~WM
  Bergmans, Eugenio Cantatore, and Massimo Mischi.
\newblock Influence of electrode placement on signal quality for ambulatory
  pregnancy monitoring.
\newblock \emph{Comput. Math. Methods Med.}, 2014, 2014{\natexlab{b}}.

\bibitem[Terrien et~al.(2010)Terrien, Steingrimsdottir, Marque, and
  Karlsson]{terrien2010synchronization}
J{\'e}r{\'e}my Terrien, Thora Steingrimsdottir, Catherine Marque, and Brynjar
  Karlsson.
\newblock Synchronization between emg at different uterine locations
  investigated using time-frequency ridge reconstruction: comparison of
  pregnancy and labor contractions.
\newblock \emph{EURASIP J. Adv. Signal Process.}, 2010:\penalty0 1--10, 2010.

\bibitem[Devedeux et~al.(1993)Devedeux, Marque, Mansour, Germain, and
  Duch{\^e}ne]{devedeux1993uterine}
Dominique Devedeux, Catherine Marque, Souheil Mansour, Guy Germain, and Jacques
  Duch{\^e}ne.
\newblock Uterine electromyography: a critical review.
\newblock \emph{Am. J. Obstet. Gynecol.}, 169\penalty0 (6):\penalty0
  1636--1653, 1993.

\bibitem[Jager et~al.(2018)Jager, Liben{\v{s}}ek, and
  Ger{\v{s}}ak]{jager2018characterization}
{[dataset]}Franc Jager, Sonja Liben{\v{s}}ek, and Ksenija Ger{\v{s}}ak.
\newblock Characterization and automatic classification of preterm and term
  uterine records.
\newblock \emph{PLoS One}, 13\penalty0 (8):\penalty0 e0202125, 2018.
\newblock \doi{https://doi.org/10.13026/C2166R}.

\bibitem[Alexandersson et~al.(2015)Alexandersson, Steingrimsdottir, Terrien,
  Marque, and Karlsson]{alexandersson2015icelandic}
{[dataset]}Asgeir Alexandersson, Thora Steingrimsdottir, Jeremy Terrien,
  Catherine Marque, and Brynjar Karlsson.
\newblock The icelandic 16-electrode electrohysterogram database.
\newblock \emph{Sci. Data}, 2\penalty0 (1):\penalty0 150017, 2015.
\newblock \doi{https://doi.org/10.1038/sdata.2015.17}.

\bibitem[Leman and Marque(2000)]{leman2000rejection}
Helene Leman and Catherine Marque.
\newblock Rejection of the maternal electrocardiogram in the electrohysterogram
  signal.
\newblock \emph{IEEE. Trans. Biomed. Eng.}, 47\penalty0 (8):\penalty0
  1010--1017, 2000.

\bibitem[Limem and Hamdi(2015)]{limem2015uterine}
Manel Limem and Mohamed~Ali Hamdi.
\newblock Uterine electromyography signals denoising using discrete wavelet
  transform.
\newblock In \emph{2015 International Conference on Advances in Biomedical
  Engineering (ICABME)}, pages 101--103. IEEE, 2015.

\bibitem[Taralunga et~al.(2015)Taralunga, Ungureanu, Hurezeanu, Gussi, and
  Strungaru]{taralunga2015empirical}
Dragos~Daniel Taralunga, Mihaela Ungureanu, Bogdan Hurezeanu, Ilinca Gussi, and
  Rodica Strungaru.
\newblock Empirical mode decomposition applied for non-invasive
  electrohysterograhic signals denoising.
\newblock In \emph{2015 37th Conf. Proc. IEEE Eng. Med. Biol. Soc. (EMBC)},
  pages 4134--4137. IEEE, 2015.

\bibitem[Huang et~al.(1998)Huang, Shen, Long, Wu, Shih, Zheng, Yen, Tung, and
  Liu]{huang1998empirical}
Norden~E Huang, Zheng Shen, Steven~R Long, Manli~C Wu, Hsing~H Shih, Quanan
  Zheng, Nai-Chyuan Yen, Chi~Chao Tung, and Henry~H Liu.
\newblock The empirical mode decomposition and the hilbert spectrum for
  nonlinear and non-stationary time series analysis.
\newblock \emph{Proc. Math. Phys. Eng. Sci.}, 454\penalty0 (1971):\penalty0
  903--995, 1998.

\bibitem[Sidiropoulos et~al.(2017)Sidiropoulos, De~Lathauwer, Fu, Huang,
  Papalexakis, and Faloutsos]{sidiropoulos2017tensor}
Nicholas~D Sidiropoulos, Lieven De~Lathauwer, Xiao Fu, Kejun Huang, Evangelos~E
  Papalexakis, and Christos Faloutsos.
\newblock Tensor decomposition for signal processing and machine learning.
\newblock \emph{IEEE Trans. Signal Process.}, 65\penalty0 (13):\penalty0
  3551--3582, 2017.

\bibitem[Cong et~al.(2015)Cong, Lin, Kuang, Gong, Astikainen, and
  Ristaniemi]{cong2015tensor}
Fengyu Cong, Qiu-Hua Lin, Li-Dan Kuang, Xiao-Feng Gong, Piia Astikainen, and
  Tapani Ristaniemi.
\newblock Tensor decomposition of eeg signals: a brief review.
\newblock \emph{J. Neurosci. Methods}, 248:\penalty0 59--69, 2015.

\bibitem[Cichocki et~al.(2015)Cichocki, Mandic, De~Lathauwer, Zhou, Zhao,
  Caiafa, and Phan]{cichocki2015tensor}
Andrzej Cichocki, Danilo Mandic, Lieven De~Lathauwer, Guoxu Zhou, Qibin Zhao,
  Cesar Caiafa, and Huy~Anh Phan.
\newblock Tensor decompositions for signal processing applications: From
  two-way to multiway component analysis.
\newblock \emph{IEEE Signal Process. Mag.}, 32\penalty0 (2):\penalty0 145--163,
  2015.

\bibitem[Zahran et~al.(2019)Zahran, Diab, Chowdhury, Hedrich, Hassan, Grova,
  Yochum, Khalil, and Marque]{zahran2019performance}
Saeed Zahran, Ahmad Diab, Rasheda~Arman Chowdhury, Tanguy Hedrich, Mahmoud
  Hassan, Christophe Grova, Maxime Yochum, Mohamad Khalil, and Catherine
  Marque.
\newblock Performance of source imaging techniques of spatially extended
  generators of uterine activity.
\newblock \emph{Inform. Med. Unlocked}, 16:\penalty0 100167, 2019.

\bibitem[Tang et~al.(2018)Tang, Chen, Wang, Zomaya, Chen, and
  Liu]{tang2018bayesian}
Yunbo Tang, Dan Chen, Lizhe Wang, Albert~Y Zomaya, Jingying Chen, and Honghai
  Liu.
\newblock Bayesian tensor factorization for multi-way analysis of
  multi-dimensional eeg.
\newblock \emph{Neurocomputing}, 318:\penalty0 162--174, 2018.

\bibitem[Jager et~al.(2020)Jager, Ger{\v{s}}ak, Vouk, Pirnar, Trojner-Bregar,
  Lu{\v{c}}ovnik, and Borovac]{jager2020assessing}
Franc Jager, Ksenija Ger{\v{s}}ak, Paula Vouk, {\v{Z}}iga Pirnar, Andreja
  Trojner-Bregar, Miha Lu{\v{c}}ovnik, and Ana Borovac.
\newblock Assessing velocity and directionality of uterine electrical activity
  for preterm birth prediction using ehg surface records.
\newblock \emph{Sensors}, 20\penalty0 (24):\penalty0 7328, 2020.

\bibitem[Duchene et~al.(1995)Duchene, Devedeux, Mansour, and
  Marque]{duchene1995analyzing}
J~Duchene, D~Devedeux, S~Mansour, and C~Marque.
\newblock Analyzing uterine emg: tracking instantaneous burst frequency.
\newblock \emph{IEEE Eng. Med. Biol. Mag.}, 14\penalty0 (2):\penalty0 125--132,
  1995.

\bibitem[Nguyen et~al.(2012)Nguyen, Peng, Do, and Liang]{nguyen2012denoising}
Hien~M Nguyen, Xi~Peng, Minh~N Do, and Zhi-Pei Liang.
\newblock Denoising mr spectroscopic imaging data with low-rank approximations.
\newblock \emph{IEEE. Trans. Biomed. Eng.}, 60\penalty0 (1):\penalty0 78--89,
  2012.

\bibitem[Zhao et~al.(2015{\natexlab{a}})Zhao, Zhou, Zhang, Cichocki, and
  Amari]{zhao2015bayesian}
Qibin Zhao, Guoxu Zhou, Liqing Zhang, Andrzej Cichocki, and Shun-Ichi Amari.
\newblock Bayesian robust tensor factorization for incomplete multiway data.
\newblock \emph{IEEE Trans. Neural. Netw. Learn. Syst.}, 27\penalty0
  (4):\penalty0 736--748, 2015{\natexlab{a}}.

\bibitem[Zhou and Cheung(2019)]{zhou2019bayesian}
Yang Zhou and Yiu-Ming Cheung.
\newblock Bayesian low-tubal-rank robust tensor factorization with multi-rank
  determination.
\newblock \emph{IEEE Trans. Pattern Anal.}, 43\penalty0 (1):\penalty0 62--76,
  2019.

\bibitem[Huang et~al.(2021)Huang, Qiu, Zhao, and Zhou]{huang2021bayesian}
Zhenhao Huang, Yuning Qiu, Qibin Zhao, and Guoxu Zhou.
\newblock Bayesian robust tucker decomposition for multiway data analysis.
\newblock In \emph{2021 China Automation Congress (CAC)}, pages 5559--5564.
  IEEE, 2021.

\bibitem[Zhao et~al.(2015{\natexlab{b}})Zhao, Zhang, and
  Cichocki]{zhao2015bayesian2}
Qibin Zhao, Liqing Zhang, and Andrzej Cichocki.
\newblock Bayesian sparse tucker models for dimension reduction and tensor
  completion.
\newblock \emph{arXiv preprint arXiv:1505.02343}, 2015{\natexlab{b}}.
\newblock \doi{https://doi.org/10.48550/arXiv.1505.02343}.

\bibitem[Diab et~al.(2021)Diab, Boudaoud, Karlsson, and
  Marque]{diab2021performance}
Ahmad Diab, Sofiane Boudaoud, Brynjar Karlsson, and Catherine Marque.
\newblock Performance comparison of coupling-evaluation methods in
  discriminating between pregnancy and labor ehg signals.
\newblock \emph{Comput. Biol. Med.}, 132:\penalty0 104308, 2021.

\bibitem[Ye-Lin et~al.(2015)Ye-Lin, Alberola-Rubio, Prats-boluda, Perales,
  Desantes, and Garcia-Casado]{ye2015feasibility}
Yiyao Ye-Lin, Jos{\'e} Alberola-Rubio, Gema Prats-boluda, A~Perales,
  D~Desantes, and Javier Garcia-Casado.
\newblock Feasibility and analysis of bipolar concentric recording of
  electrohysterogram with flexible active electrode.
\newblock \emph{Ann. Biomed. Eng.}, 43:\penalty0 968--976, 2015.

\bibitem[Rabotti et~al.(2010)Rabotti, Mischi, Oei, and
  Bergmans]{rabotti2010noninvasive}
Chiara Rabotti, Massimo Mischi, S~Guid Oei, and Jan~WM Bergmans.
\newblock Noninvasive estimation of the electrohysterographic action-potential
  conduction velocity.
\newblock \emph{IEEE. Trans. Biomed. Eng.}, 57\penalty0 (9):\penalty0
  2178--2187, 2010.

\bibitem[Mikkelsen et~al.(2013)Mikkelsen, Johansen, Fuglsang-Frederiksen, and
  Uldbjerg]{mikkelsen2013electrohysterography}
Eva Mikkelsen, Peter Johansen, Anders Fuglsang-Frederiksen, and Niels Uldbjerg.
\newblock Electrohysterography of labor contractions: propagation velocity and
  direction.
\newblock \emph{Acta Obstet. Gynecol. Scand.}, 92\penalty0 (9):\penalty0
  1070--1078, 2013.

\bibitem[Aravkin et~al.(2014)Aravkin, Becker, Cevher, and
  Olsen]{aravkin2014variational}
Aleksandr Aravkin, Stephen Becker, Volkan Cevher, and Peder Olsen.
\newblock A variational approach to stable principal component pursuit.
\newblock \emph{arXiv preprint arXiv:1406.1089}, 2014.
\newblock \doi{https://doi.org/10.48550/arXiv.1406.1089}.

\bibitem[Chambolle et~al.(2010)Chambolle, Caselles, Cremers, Novaga, and
  Pock]{chambolle2010introduction}
Antonin Chambolle, Vicent Caselles, Daniel Cremers, Matteo Novaga, and Thomas
  Pock.
\newblock An introduction to total variation for image analysis.
\newblock \emph{Theoretical Foundations and Numerical Methods for Sparse
  Recovery}, 9\penalty0 (263-340):\penalty0 227, 2010.

\bibitem[Lourakis(2017)]{L1_matlab}
Manolis Lourakis.
\newblock Tv-l1 image denoising algorithm.
\newblock
  \url{https://www.mathworks.com/matlabcentral/fileexchange/57604-tv-l1-image-denoising-algorithm},
  2017.
\newblock Accessed: 2022-3-21.

\bibitem[Hore et~al.(2016)Hore, Vi{\~n}uela, Buil, Knight, McCarthy, Small, and
  Marchini]{hore2016tensor}
Victoria Hore, Ana Vi{\~n}uela, Alfonso Buil, Julian Knight, Mark~I McCarthy,
  Kerrin Small, and Jonathan Marchini.
\newblock Tensor decomposition for multiple-tissue gene expression experiments.
\newblock \emph{Nat. Genet.}, 48\penalty0 (9):\penalty0 1094--1100, 2016.

\bibitem[Zhang et~al.(2019)Zhang, Wang, Fu, Zhong, and
  Huang]{zhang2019computational}
Shipeng Zhang, Lizhi Wang, Ying Fu, Xiaoming Zhong, and Hua Huang.
\newblock Computational hyperspectral imaging based on dimension-discriminative
  low-rank tensor recovery.
\newblock In \emph{Proceedings of the IEEE/CVF International Conference on
  Computer Vision}, pages 10183--10192, 2019.

\bibitem[Laforet et~al.(2011)Laforet, Rabotti, Terrien, Mischi, and
  Marque]{laforet2011toward}
Jeremy Laforet, Chiara Rabotti, Jeremy Terrien, Massimo Mischi, and Catherine
  Marque.
\newblock Toward a multiscale model of the uterine electrical activity.
\newblock \emph{IEEE. Trans. Biomed. Eng.}, 58\penalty0 (12):\penalty0
  3487--3490, 2011.

\bibitem[La~Rosa et~al.(2012)La~Rosa, Eswaran, Preissl, and
  Nehorai]{la2012multiscale}
Patricio~S La~Rosa, Hari Eswaran, Hubert Preissl, and Arye Nehorai.
\newblock Multiscale forward electromagnetic model of uterine contractions
  during pregnancy.
\newblock \emph{BMC Med. Phys.}, 12\penalty0 (1):\penalty0 1--16, 2012.

\bibitem[Zhang et~al.(2016)Zhang, Tidwell, La~Rosa, Wilson, Eswaran, and
  Nehorai]{zhang2016modeling}
Mengxue Zhang, Vanessa Tidwell, Patricio~S La~Rosa, James~D Wilson, Hari
  Eswaran, and Arye Nehorai.
\newblock Modeling magnetomyograms of uterine contractions during pregnancy
  using a multiscale forward electromagnetic approach.
\newblock \emph{PLoS One}, 11\penalty0 (3):\penalty0 e0152421, 2016.

\bibitem[Esgalhado et~al.(2020)Esgalhado, Batista, Mouri{\~n}o, Russo,
  Dos~Reis, Serrano, Vassilenko, and Ortigueira]{esgalhado2020uterine}
Filipa Esgalhado, Arnaldo~G Batista, Helena Mouri{\~n}o, Sara Russo, Catarina
  R~Palma Dos~Reis, F{\'a}tima Serrano, Valentina Vassilenko, and Manuel
  Ortigueira.
\newblock Uterine contractions clustering based on electrohysterography.
\newblock \emph{Comput. Biol. Med.}, 123:\penalty0 103897, 2020.

\end{thebibliography}

\end{document}